\documentclass[showpacs,amssymb,preprintnumbers,nofootinbib,
superscriptaddress]{revtex4}

\usepackage{epsfig}
\usepackage{amsmath}
\usepackage{graphicx}
\usepackage{latexsym}
\usepackage{amsfonts}
\usepackage{url,hyperref}
\usepackage{bm}
\usepackage{textcomp}
\usepackage{color}
\newcommand{\nn}{\nonumber}
\newcommand{\beq}{\begin{equation}}
\newcommand{\eeq}{\end{equation}}
\def\bea{\begin{eqnarray}}
\def\eea{\end{eqnarray}}

\begin{document}

\title{Evolution of the CKM Matrix in the Universal Extra Dimension Model}
\author{A.~S.~Cornell}
\email[Email: ]{alan.cornell@wits.ac.za}
\affiliation{National Institute for Theoretical Physics; School of Physics, University of the Witwatersrand, Wits 2050, South Africa}
\author{Lu-Xin~Liu}
\email[Email: ]{luxin.liu9@gmail.com}
\affiliation{National Institute for Theoretical Physics; School of Physics, University of the Witwatersrand, Wits 2050, South Africa}

\begin{abstract}
The evolution of the Cabibbo-Kobayashi-Maskawa matrix and the quark Yukawa couplings is performed for the one-loop renormalization group equations in the universal extra dimension model. It is found that the evolution of mixing angles and the CP violation measure $J$ may rapidly vary in the presence of the Kaluza-Klein modes, and this variation becomes dramatic as the energy approaches the unification scale. 
\end{abstract}

\date{October 26, 2010}
\preprint{WITS-CTP-58}
\maketitle

%%%%%%%%%%%%%%%%%%%%%%%%%%%%%%%%%

\section{Introduction}\label{sec:1}

\par With the Large Hadron Collider (LHC) now up and running, exploration of the realm of new physics that may operate at the TeV scale has begun. Among these models those with extra spatial dimensions might be revealed in such higher energy collider experiments, where the universal extra dimension (UED) model makes for an interesting TeV scale physics scenario; as it features a tower of Kaluza-Klein (KK) states for each of the standard model (SM) fields, all of which have full access to the extended spacetime manifold \cite{Appelquist:2000nn}. This particular scenario has recently been extensively studied in the literature, such as investigations of electroweak symmetry breaking, proton stability, gauge hierarchy and fermion mass hierarchy problems, $B$ physics, dark matter etc. \cite{Arkani-Hamed:2000, Appelquist:2001nn, Bhattacharyya:2006ym, Buras:2003mk, Hooper:2007qk, Datta:2010us}. This model has been a fruitful playground for addressing a variety of puzzles in the SM.  

\par On the other hand, it is well known that in the SM, the quark sector's flavor mixing is parameterized by the Cabibbo-Kobayashi-Maskawa (CKM) matrix:
\beq
V_{CKM} = \left( {\begin{array}{ccc}
{{V_{ud}}}&{{V_{us}}}&{{V_{ub}}}\\
{{V_{cd}}}&{{V_{cs}}}&{{V_{cb}}}\\
{{V_{td}}}&{{V_{ts}}}&{{V_{tb}}}
\end{array}} \right) = \left( {\begin{array}{ccc}
{{V_{11}}}&{{V_{12}}}&{{V_{13}}}\\
{{V_{21}}}&{{V_{22}}}&{{V_{23}}}\\
{{V_{31}}}&{{V_{32}}}&{{V_{33}}}
\end{array}} \right) \; , \label{eqn:1}
\eeq
which makes it possible to explain all flavor changing weak decay process and CP-violating phenomena to date, where the 10 year run of Babar at SLAC and the Belle detector at KEK has greatly improved our knowledge of the CKM matrix elements. In particular, for the standard parameterization of the CKM matrix, which has the form: 
\beq
V_{CKM} = \left( {\begin{array}{ccc}
{{c_{12}}{c_{13}}}&{{s_{12}}{c_{13}}}&{{s_{13}}{e^{ - i{\delta}}}}\\
{ - {s_{12}}{c_{23}} - {c_{12}}{s_{23}}{s_{13}}{e^{i{\delta}}}}&{{c_{12}}{c_{23}} - {s_{12}}{s_{23}}{s_{13}}{e^{i{\delta}}}}&{{s_{23}}{c_{13}}}\\
{{s_{12}}{s_{23}} - {c_{12}}{c_{23}}{s_{13}}{e^{i{\delta}}}}&{ - {c_{12}}{s_{23}} - {s_{12}}{c_{23}}{s_{13}}{e^{i{\delta}}}}&{{c_{23}}{c_{13}}}
\end{array}} \right) \; , \label{eqn:2}
\eeq
where $s_{12} = \sin\theta_{12}$, $c_{12} = \cos\theta_{12}$ etc. are the sines and cosines of the three mixing angles $\theta_{12}$, $\theta_{23}$ and $\theta_{13}$, and $\delta$ is the CP violating phase.

\par The CKM matrix highlights that, in the quark sector of the SM, we have ten experimentally measurable parameters, i.e. six quark masses, three mixing angles, and one phase. A completely satisfactory theory of fermion masses and the related problem of mixing angles is certainly lacking at present, however, there has been considerable effort to understand the hierarchies of these mixing angles and fermion masses in terms of the renormalization group equations (RGE) \cite{Cheng:1973nv, Babu:1987im, Sasaki:1986jv, Machacek:1983fi, Liu:2009vh, Balzereit:1998id, Kuo:2005jt}.  Recall that in order to explore the physics at a high energy scale we use RGE as a probe to study the momentum dependence of the Yukawa couplings, the gauge couplings, and the CKM matrix elements themselves. As such we can consider one of the primary goals of the LHC as being to uncover any new dynamics within the TeV range. Instead of assuming the RGE goes from the $M_Z$ scale up to the GUT scale ($10^{15}$ GeV) by using the $SU_C(3)\times SU_L(2) \times U_Y(1)$ symmetry, we know that models with extra dimensions may bring down the unification to a much lower energy scale. 

\par The UED model we shall consider places all SM particles in the bulk of one or more compactified extra dimensions. In the simplest case, there is a single flat extra dimension of size $R$, compactified on an $S_1/Z_2$ orbifold. Therefore, from a 4-dimensional view point, every field will have an infinite tower of KK modes, with the zero modes being identified as the SM states. If these KK modes are indeed within the TeV range, they would modify the running of the RGE at relative low energy scales. However, like any higher dimensional theory, the UED model should be treated only as an effective theory which is valid up to some scale $\Lambda$, at which a new physics theory emerges. Between the scale $R^{-1}$ where the first KK states are excited and the cutoff scale $\Lambda$, there are finite quantum corrections to the Yukawa and gauge couplings from the $\Lambda R$ number of KK states. Up to the scale $R^{-1}$ the first step KK excitation occurs, the RG evolution is logarithmic, controlled by the SM beta functions. With the increasing of the energy, that is, when the KK threshold is crossed for each successive mode, new excitations come into play and govern new sets of beta functions. The values of physical parameters such as Yukawa couplings and gauge couplings do not run in the old SM fashion, instead they receive finite quantum corrections whose magnitudes depend explicitly on the value of this cutoff parameter.  As a result, once the KK states are excited, these couplings exhibit power law dependences on $\Lambda$. This can be illustrated if $\Lambda R \gg 1$, to a very good accuracy, the generic SM beta function is shown to have the power law evolution behavior \cite{Bhattacharyya:2006ym}
\beq
{\beta ^{SM}} \to {\beta ^{SM}} + (S(\mu ) - 1) {\tilde \beta } \; , \label{eqn:3}
\eeq
where $\tilde{\beta}$ is a generic contribution from a single KK level, and where its coefficient is not a constant but instead $S(\mu )=\mu R$, with $\mu^{Max}=\Lambda$, reflecting the power law running behavior. As a result of faster running, the gauge couplings tend to lower the unification scale down to a relatively low order, which might be accessible to collider experiments, such as the Tevatron Run-II, LHC or the proposed international linear collider (ILC).  Therefore, constraints from precision electroweak tests and current (or future) collider data would yield bounds on the compactification radius $R$. 

\par The RGE are an important tool for the search of the properties of the quark masses and the CKM matrix at different energy scales. It is therefore of great interest to have an implementation of the UED model in studying these RGE. In this paper, we consider the UED model with a single compactified extra dimension. In section 2, we start from the KK expansion of the SM fields and show the contributions of the relevant one-loop diagrams to the Yukawa couplings at each KK level. The RGE of the Yukawa couplings are then derived by using the anomalous dimensions of the wave function renormalization of the relevant fields and the vertex renormalization. We shall then derive the one-loop RGE for the CKM matrix.  In section 3 we shall quantitatively analyze the evolution of these RGE from low energies up to the unification scale. The scale dependence of the mixing angles as well as the CP violation measure $J$ will be plotted. We also calculate their evolution behaviors for different compactification radii $R$. The last section is devoted to a summary and our conclusions.  

%%%%%%%%%%%%%%%%%%%%%%%%%%%%%%%%%%

\section{Renormalization Group Equations}\label{sec:2}

\par
\begin{figure}[tb]
\begin{center}
\epsfig{file=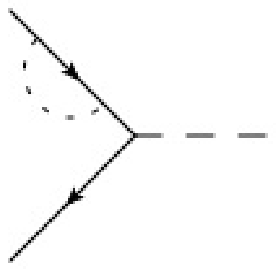,width=.25\textwidth}
\epsfig{file=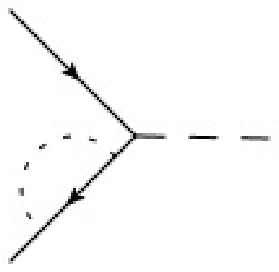,width=.25\textwidth}
\epsfig{file=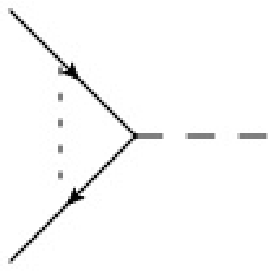,width=.25\textwidth}
\caption{\sl The one-loop corrections of the additional diagrams (to the SM type diagrams) from the fifth component of the vector fields to the Yukawa couplings in Eq.(\ref{eqn:5}), introduced at each KK excited level. The dashed line is for the Higgs field, the dotted line is for the $A_5$ scalar. In Table \ref{tab:1} the contributions to the anomalous dimension for each these possible diagrams is presented.} 
\label{fig:1}
\end{center}
\end{figure}

 Starting with the SM, which is based on the group structure $SU_C(3)\times SU_L(2) \times U_Y(1)$ with one Higgs doublet, the mass matrices arise from the Yukawa sector of the theory as given by
\beq
{\cal L}_Y = {Y_U}\varepsilon H\bar uQ + {Y_D}{H^*}\bar dQ + {Y_E}{H^*}\bar eL \; , \label{eqn:4}
\eeq
where $Q$ and $L$ are the $SU_L(2)$ doublets for the quark and lepton sectors, $u$, $d$ and $e$ are the right-handed $SU_L(2)$ singlets for up-type, down-type quarks and leptons respectively, $H$ is the Higgs doublet, and $\varepsilon$ is the $2\times 2$ antisymmetric tensor, with $\varepsilon_{12} = - \varepsilon_{21} = - 1$. The above equation implicitly contains the summation of the quark generations as well as the summation over the $SU_L(2)$ indices. The information of the physical observables at the scale $M_Z$ can be extrapolated to a higher energy scale by means of the RGE. It is well known that the evolution of the generic Yukawa coupling  
\beq
Y{{\bar \psi }_R}{\psi _L}\phi \; , \label{eqn:5}
\eeq
which describes the fermion-boson interactions, is given by the beta function. Although the bare constants are independent of the renormalization scale, the renormalized coupling constants will depend on the choice of the scale parameter $\mu$.  As a result, the Yukawa coupling renormalization depends on the corresponding beta functions, including contributions from the anomalous dimensions of the field operators. That is, its evolution is given by: 
\beq
\mu \frac{\partial }{{\partial \mu }}\ln {Y^R} = \frac{1}{2}\mu \frac{\partial }{{\partial \mu }}\ln {Z_{{\psi _L}}} + \frac{1}{2}\mu \frac{\partial }{{\partial \mu }}\ln {Z_{{\psi _R}}} + \frac{1}{2}\mu \frac{\partial }{{\partial \mu }}\ln {Z_\phi } - \mu \frac{\partial }{{\partial \mu }}\ln {Z_{couping}} \; , \label{eqn:6}
\eeq
where $Y^R$ is the renormalized Yukawa coupling constant (we shall drop the index $R$ for the remainder of the paper), and $Z_{\psi_L}$, $Z_{\psi_R}$ and $Z_\phi$ are the wave function renormalization constants related to left-handed, right-handed fermions and Higgs boson respectively, and $Z_{coupling}$ is the vertex renormalization constant; or in terms of the anomalous dimensions, $\displaystyle \gamma_{wave} = \frac{1}{2} \mu \frac{\partial}{\partial \mu} \mathrm{ln} Z_{wave}, \displaystyle \gamma_{coupling} = \mu \frac{\partial}{\partial \mu} \mathrm{ln} Z_{coupling}$. In the current context, we have chosen to work with the minimal UED model, i.e. the extra dimension is compactified on a circle of radius $R$ with a $Z_2$ orbifolding, which identifies the fifth coordinate $y \to -y$. The 5-dimensional KK expansions of the weak doublet and singlet as well as the Higgs field are shown (the corresponding coupling constants among the KK modes are simply equal to the SM couplings up to normalization factors, e.g. $\displaystyle Y_U = \frac{Y_U^5}{\sqrt{\pi R}}$) as below: 
\bea
H(x,y) &=& \frac{1}{{\sqrt {\pi R} }}\left\{ H(x) + \sqrt 2 \sum\limits_{n = 1}^ \propto  {{H_n}(x)\cos \left(\frac{{ny}}{R}\right)} \right\} \; , \nn \\
u(x,y) &=& \frac{1}{{\sqrt {\pi R} }}\left\{ {u_R}(x) + \sqrt 2 \sum\limits_{n = 1}^\infty  \left[u_R^n (x)\cos \left(\frac{{ny}}{R}\right) + u_L^n(x)\sin \left(\frac{{ny}}{R}\right)\right]\right\} \; , \nn \\
Q(x,y) &=& \frac{1}{{\sqrt {\pi R} }}\left\{ {q_L}(x) + \sqrt 2 \sum\limits_{n = 1}^\infty  \left[Q_L^n (x)\cos \left(\frac{{ny}}{R}\right) + Q_R^n(x)\sin \left(\frac{{ny}}{R}\right)\right]\right\} \; , \nn \\
d(x,y) &=& \frac{1}{{\sqrt {\pi R} }}\left\{ {d_R}(x) + \sqrt 2 \sum\limits_{n = 1}^\infty  \left[d_R^n(x)\cos \left(\frac{{ny}}{R}\right) + d_L^n(x)\sin \left(\frac{{ny}}{R}\right)\right]\right\} \; , \nn\\
L(x,y) &=& \frac{1}{{\sqrt {\pi R} }}\left\{ {L_L}(x) + \sqrt 2 \sum\limits_{n = 1}^\infty  \left[L_L^n (x)\cos \left(\frac{{ny}}{R}\right) + L_R^n(x)\sin \left(\frac{{ny}}{R}\right)\right]\right\} \; , \nn\\
e(x,y) &=& \frac{1}{{\sqrt {\pi R} }}\left\{ {e_R}(x) + \sqrt 2 \sum\limits_{n = 1}^\infty  \left[e_R^n (x)\cos \left(\frac{{ny}}{R}\right) + e_L^n(x)\sin \left(\frac{{ny}}{R}\right)\right]\right\} \; . \label{eqn:7}
\eea
The zero modes in the above equations are identified with the 4-dimensional SM fields, whilst the complex scalar field $H$ is $Z_2$ even field, and there is a left-handed and a right-handed KK mode for each SM chiral fermion. Note that in models with UED momentum conservation in the extra dimensions, we are led to the conservation of KK number at each vertex in the interactions of the 4-dimensional effective theory (or strictly speaking, the KK parity $(-1)^n$ is what remains conserved, where $n$ is the KK number). In the bulk we have the fermion and gauge field interactions as follows:
\bea
{\cal L}_{Leptons} &=& \int\limits_0^{\pi R} {dy} \{ i\bar L(x,y){\Gamma ^M}{{\cal D}_M}L(x,y) + i\bar e(x,y){\Gamma ^M}{{\cal D}_M}e(x,y)\} \; , \nn \\
{\cal L}_{Quarks} &=& \int\limits_0^{\pi R} {dy} \{ i\bar Q(x,y){\Gamma ^M}{{\cal D}_M}Q(x,y) + i\bar u(x,y){\Gamma ^M}{{\cal D}_M}u(x,y) + i\bar d(x,y){\Gamma ^M}{{\cal D}_M}d(x,y)\} \; , \label{eqn:8}
\eea
where $\Gamma^M=(\gamma^{\mu},i\gamma^5)$, and $M=0,1,2,3,5$. Explicitly, the kinetic terms are given by:
\bea
{{\cal D}_M}Q(x,y) &=& \left({\partial _M} + ig_3^5{G_M} + ig_2^5{W_M} + i\frac{{{1}}}{6}g_1^5{B_M}\right)Q(x,y)\; , \nn\\
{{\cal D}_M}u(x,y) &=& \left({\partial _M} + ig_3^5{G_M} + i\frac{{{2}}}{3}g_1^5{B_M}\right)u(x,y)\; , \nn\\
{{\cal D}_M}d(x,y) &=& \left({\partial _M} + ig_3^5{G_M} + i\frac{{{-1}}}{3}g_1^5{B_M}\right)d(x,y)\; , \nn\\
{{\cal D}_M}L(x,y) &=& \left({\partial _M} + ig_2^5{W_M} + i\frac{{{-1}}}{2}g_1^5{B_M}\right)L(x,y)\; , \nn\\
{{\cal D}_M}e(x,y) &=& \left({\partial _M} - ig_1^5{B_M}\right)e(x,y) \; . \label{eqn:9}
\eea
The gauge couplings $g_3^5$, $g_2^5$ and $g_1^5$ refer to those of the $SU(3)$, $SU(2)$ and $U(1)$ gauge groups respectively, and are related to the 4-dimensional SM coupling constants $\displaystyle g_i = \frac{g_i^5}{\sqrt{\pi R}}$. After integrating out the compactified dimension, the 4-dimensional effective Lagrangian has interactions involving the zero mode and the KK modes. However, these KK modes cannot affect electroweak process at tree level, and only contribute to higher order electroweak processes. For the one-loop diagrams of the Yukawa couplings, we choose the Landau gauge in what follows, as many one-loop diagrams are finite in the Landau gauge and have no contribution to the renormalization of the Yukawa couplings. We therefore consider the RGE for the quark-Higgs Yukawa couplings from which we obtain the evolution of the quark masses and the CKM matrix. The one-loop Feynman diagram contributions to the Yukawa couplings in the SM and UED model have been explicitly illustrated in Ref.\cite{Bhattacharyya:2006ym, Cheng:1973nv}. In the UED model, where for each energy level $n_i$, we effectively have a heavier duplicate copy of the entire SM particle content.  That is, at each KK excited level, the KK tower corresponding to the fields in Eq.(\ref{eqn:7}) exactly mirror the SM field ground states. However, new contributions from the $A_5$,
\beq
A_5(x,y) = \sqrt{\frac{2}{\pi R}} \sum\limits_{n = 1}^\infty A_5^n(x)\sin\left(\frac{ny}{R}\right)  \; , \label{eqn:10}
\eeq
interactions (that of the fifth component of the vector fields, i.e. $A_5=\{G_5, W_5, B_5\}$), as illustrated in Fig.\ref{fig:1}, also contribute. In contrast, the fifth component of the gauge bosons $A_5 (x,y)$ is a real scalar and does not have any zero mode, transforming in the adjoint representation of the gauge group. 

%%%%%%%%%%%%%%%%
%%%%%%%%%%%%%%%%
%%%%%%%%%%%%%%%%
\begin{table}[t]
\caption{\small\sl The anomalous dimensions for each diagram in Fig.\ref{fig:1}. Note that the columns refer to each type of fifth component for the vector fields ($g_1$, $g_2$ and $g_3$), for each type of wavefunction renormalization and proper Yukawa vertex renormalization, along with their corresponding contributions in the rows for each type of Yukawa beta function.}
\label{tab:1}
\begin{ruledtabular}
\begin{tabular}{|c|ccc|ccc|ccc|}
 & & $\gamma _{{\psi _L}}$ & & & $\gamma _{{\psi _R}}$ & & & $\gamma _{Coupling}$ &  \\ \hline
 Yukawa beta function &$g_1$&$g_2$&$g_3$&$g_1$&$g_2$&$g_3$&$g_1$&$g_2$&$g_3$ \\ \hline 
$Up-type\; quark$ &$\frac{1}{2}{g_1}^2{(\frac{1}{6})^2\frac{3}{5}}$&$\frac{1}{2}{g_2}^2\frac{3}{4}$&$\frac{1}{2}{g_3}^2\frac{8}{6}$&$\frac{1}{2}{g_1}^2{(\frac{2}{3})^2}\frac{3}{5}$&&$\frac{1}{2}{g_3}^2\frac{8}{6}$&${g_1}^2\frac{2}{9}\frac{3}{5}$& &$ {g_3}^2\frac{8}{3}$ \\ 
$Down-type\; quark$ &$\frac{1}{2}{g_1}^2{(\frac{1}{6})^2}\frac{3}{5}$&$\frac{1}{2}{g_2}^2\frac{3}{4}$&$\frac{1}{2}{g_3}^2\frac{8}{6}$&$\frac{1}{2}{g_1}^2{( - \frac{1}{3})^2}\frac{3}{5}$&&$\frac{1}{2}{g_3}^2\frac{8}{6}$&$-{g_1}^2\frac{1}{9}\frac{3}{5}$&&${g_3}^2\frac{8}{3}$\\ 
$Lepton\; sector$ &$\frac{1}{2}{g_1}^2\frac{3}{5}$&${g_2}^2\frac{3}{8}$&&$\frac{1}{8}{g_1}^2\frac{3}{5}$&&&${g_1}^2\frac{3}{5}$& & \end{tabular}
\end{ruledtabular}
\end{table}
%%%%%%%%%%%%%%%%

\par That is, each of the other graphs which contribute, due to the new excitations, exactly mirror the zero mode SM ground state, where their contributions to the anomalous dimensions are exactly the same as those in the usual 4-dimensional SM. However, there is a subtlety here, due to the existence of both left- and right-handed chiral KK modes in Eq.(\ref{eqn:7}) (when the expansion of fermion fields is done), that is, we need to double count the anomalous dimension contributions in contrast with those of the SM zero modes. We use dimensional regularization to extract the anomalous dimensions from the divergent parts of the above one-loop diagrams, where in Table \ref{tab:1} we list the results of the anomalous dimensions for the relevant fields and vertices in Fig.\ref{fig:1}, related to the up-type, down-type quarks, and lepton Yukawa couplings respectively; where for simplicity we have omitted a common multiplicative factor of $\displaystyle \frac{1}{16 \pi^2}$ in the table. The one-loop RGE for the Yukawa couplings take the following form:
\bea
16{\pi ^2}\frac{{d{Y_U}}}{{dt}} &= & \beta _U^{SM} + \beta _U^{UED} \; , \nn \\
16{\pi ^2}\frac{{d{Y_D}}}{{dt}} &=& \beta _D^{SM} + \beta _D^{UED} \; , \label{eqn:11}
\eea
where the Yukawa coupling beta functions have the form:                                  
\bea
\beta _U^{SM} &=& {Y_U}\left\{  - \left(8g_3^2 + \frac{9}{4}g_2^2 + \frac{{17}}{{20}}g_1^2\right) + \frac{3}{2}\left(Y_U^\dag {Y_U} - Y_D^\dag {Y_D}\right) + Tr\left[3Y_U^\dag {Y_U} + 3Y_D^\dag {Y_D} + Y_E^\dag {Y_E}\right]\right\} \; , \nn \\
\beta _U^{UED} &=& {Y_U}\left\{ (S(t) - 1)\left[ - \left(\frac{{28}}{3}g_3^2 + \frac{{15}}{8}g_2^2 + \frac{{101}}{{120}}g_1^2\right) + \frac{3}{2}\left(Y_U^\dag {Y_U} - Y_D^\dag {Y_D}\right) \right] \right. \nn\\
&& \hspace{1cm} \left. + 2(S(t) - 1)Tr\left[3Y_U^\dag {Y_U} + 3Y_D^\dag {Y_D} + Y_E^\dag {Y_E}\right]\right\} \; , \label{eqn:12}
\eea
and
\bea
\beta _D^{SM} &=& {Y_D}\left\{  - \left(8g_3^2 + \frac{9}{4}g_2^2 + \frac{1}{4}g_1^2\right) + \frac{3}{2}\left(Y_D^\dag {Y_D} - Y_U^\dag {Y_U}\right) + Tr\left[3Y_U^\dag {Y_U} + 3Y_D^\dag {Y_D} + Y_E^\dag {Y_E}\right]\right\} \; , \nn \\
\beta _D^{UED} &=& {Y_D}\left\{ (S(t) - 1)\left[ - \left(\frac{{28}}{3}g_3^2 + \frac{{15}}{8}g_2^2 + \frac{{17}}{{120}}g_1^2\right) + \frac{3}{2}\left(Y_D^\dag {Y_D} - Y_U^\dag {Y_U}\right)\right] \right. \nn \\
&& \left. \hspace{1cm} + 2(S(t) - 1)Tr\left[3Y_U^\dag {Y_U} + 3Y_D^\dag {Y_D} + Y_E^\dag {Y_E}\right]\right\} \; . \label{eqn:13}
\eea
Note that here $t = \mathrm{ln} (\mu/M_Z)$ is the energy scale parameter, and $S(t)=e^t M_Z R$, where we have chosen the $Z$ boson mass as the renormalization point. The coupling constant $g_1$ is also chosen to follow the conventional $SU(5)$ normalization. 

\par Note that in deriving Eqs.(\ref{eqn:12},\ref{eqn:13}), for the factor of $g_3^2$ in the UED beta function of Eq.(\ref{eqn:12}), as an explicit example, we have $\gamma _{{\psi _L}}+\gamma _{{\psi _R}}-\gamma _{Coupling}=-4/3 g_3^2$ from Table \ref{tab:1}. Together with a factor $-8 g_3^2$ which is read off from the first line of Eq.(\ref{eqn:12}). This results from other one-loop graphs of the KK modes which mirror those of the SM zero mode. This then leads to a total factor $-28 g_3^2/3$. At this point, our results differ from those of Ref. \cite{Bhattacharyya:2006ym}. In Table \ref{tab:1} we have positive signs for the wave function anomalous dimensions.  Conventionally, the sign of $\gamma_{wave}$ can be justified by considering the fact that the wave function renormalization constant is less than unity for the gauge independent  interacting field \cite{Ticciati:2010zzi}, and the divergent parts of these one-loop diagrams are independent of $A_5$ gauge dependencies. Note that while the zero mode fermions are chiral as a result of orbifolding, the KK quarks and leptons at given levels are vector like. This accounts for the relative factor of 2 between the first and second terms of the UED beta function in Eqs.(\ref{eqn:12},\ref{eqn:13}) for the overall proportionality factor $S-1$, since both the KK left- and right-handed chiral states can simultaneously contribute to the closed fermion one-loop diagrams. The interaction between $A_5$ and the Higgs field only contributes to the renormalization of the Higgs mass, thus the wave function anomalous dimensions of the Higgs field is immune to the effects of $A_5$.  Similarly, for the lepton sector we have:
\beq
16{\pi ^2}\frac{{d{Y_E}}}{{dt}} = \beta _E^{SM} + \beta _E^{UED} \; , \label{eqn:14}
\eeq
in which
\bea
\beta_E^{SM} &=& {Y_E}\left\{ - \left(\frac{9}{4}g_2^2 + \frac{9}{4}g_1^2\right) + \frac{3}{2}Y_E^\dag {Y_E}+Tr\left[3Y_U^\dag {Y_U} + 3Y_D^\dag {Y_D} + Y_E^\dag {Y_E}\right] \right\} \; , \nn \\
\beta _E^{UED} &=& {Y_E}\left\{ (S(t) - 1)\left[ - \left(\frac{{15}}{8}g_2^2 + \frac{{99}}{{40}}g_1^2\right) + \frac{3}{2}Y_E^\dag {Y_E}\right] + 2(S(t) - 1)Tr\left[3Y_U^\dag {Y_U} + 3Y_D^\dag {Y_D} + Y_E^\dag {Y_E}\right]\right\} \; . \label{eqn:15}
\eea
Or, explicitly in diagonal form
\beq
16{\pi ^2}\frac{{d{y_a}^2}}{{dt}} = {y_a}^2\left[2(2S(t) - 1)T - 2{G_E} + 3S(t){y_a}^2\right] \; , \label{eqn:16}
\eeq
where we use $Y_E = \mathrm{diag}(y_e, y_\mu, y_\tau)$, $G_E = \displaystyle \left( \frac{9}{4} g_2^2 + \frac{9}{4} g_1^2 \right) + (S(t)-1) \left( \frac{15}{8} g_2^2 + \frac{99}{40} g_1^2 \right)$, and $T=Tr[3Y_U^\dag {Y_U} + 3Y_D^\dag {Y_D} + Y_E^\dag {Y_E}]$. Eqs.(\ref{eqn:11},\ref{eqn:16}) constitute a complete set of coupled differential equations for the three families.     

\par Further, after diagonalizing the square of the quark Yukawa coupling matrices by using two unitary matrices $U$ and $V$,  
\bea
UY_U^\dag {Y_U}{U^\dag } &=& \mathrm{diag}\left(f_u^2,f_c^2,f_t^2\right) \; , \nn \\
VY_D^\dag {Y_D}{V^\dag } &=& \mathrm{diag}\left(h_d^2,h_s^2,h_b^2\right) \; , \label{eqn:17}
\eea
in which $f_u^2$, $f_c^2$, $f_t^2$ and $h_d^2$, $h_s^2$, $h_b^2$ are the eigenvalues of $Y_U^\dag Y_U$ and $ Y_D^\dag Y_D$ respectively, it follows that the CKM matrix describing the quark flavor mixing in the charged current is given by 
\beq
{V_{CKM}} = U{V^\dag }\; . \label{eqn:18}
\eeq
We will now obtain from Eq.(\ref{eqn:11}) the RGE of the elements of the CKM matrix. Note that we can see from these equations that one needs to know the running of Eq.(\ref{eqn:11}) to obtain the evolution of the CKM matrix. By imposing the unitary transformation, Eq.(\ref{eqn:17}), on both sides of the evolution equations of $Y_U^\dag Y_U$ and $Y_D^\dag Y_D$, and taking the diagonal elements, we obtain the following two relations: 
\bea 
16{\pi ^2}\frac{{d{f_i}^2}}{{dt}} &=& {f_i}^2\left[2(2S(t) - 1)T - 2{G_U} + 3S(t){f_i}^2 - 3S(t)\sum\limits_j  {{h_j }^2} {\left| {{V_{ij}}} \right|^2}\right] \; , \nn\\
16{\pi ^2}\frac{{d{h_j }^2}}{{dt}} &=& {h_j }^2 \left [2(2S(t) - 1)T - 2{G_D} + 3S(t){h_j }^2 - 3S(t)\sum\limits_i {{f_i}^2} {\left| {{V_{ij }}} \right|^2}\right]\; , \label{eqn:19}
\eea
which describe the evolution equations of $f_u^2$, $f_c^2$, $f_t^2$ and $h_d^2$, $h_s^2$, $h_b^2$ respectively, where                           
\bea
{G_U} &=& 8g_3^2 + \frac{9}{4}g_2^2 + \frac{{17}}{{20}}g_1^2 + (S(t) - 1)\left(\frac{{28}}{3}g_3^2 + \frac{{15}}{8}g_2^2 + \frac{{101}}{{120}}g_1^2\right) \; , \nn\\
{G_D} &=& 8g_3^2 + \frac{9}{4}g_2^2 + \frac{1}{4}g_1^2 + (S(t) - 1)\left(\frac{{28}}{3}g_3^2 + \frac{{15}}{8}g_2^2 + \frac{{17}}{{120}}g_1^2\right)\; . \label{eqn:20}
\eea
Recall that the eigenvalues of the Yukawa coupling matrices are the ratios of the fermion masses to the Higgs vacuum exception value $v$, i.e. $\displaystyle m = \frac{{v}}{\sqrt{2}} Y$ (where $v=246$ GeV). The quark mass matrices then appear after the spontaneous symmetry breaking from the quark-Higgs Yukawa couplings, where Eq.(\ref{eqn:19}) describes how the squares of the physical Yukawa couplings evolve. Considering the variation of the square of the Yukawa couplings, we may impose two new unitary matrices to make them diagonal. Thus, by applying Eq.(\ref{eqn:18}), we are led to the variation of the CKM matrix and thus its evolution equation (beyond the threshold $R^{-1}$):        
\beq
16{\pi ^2}\frac{{d{V_{ik }}}}{{dt}} =  - \frac{3}{2}S(t)\left[\sum\limits_{m ,j \ne i} {\frac{{{f_i}^2 + {f_j}^2}}{{{f_i}^2 - {f_j}^2}}} {h_m }^2{V_{im }}{V_{jm }}^*{V_{jk }} + \sum\limits_{j,m  \ne k } {\frac{{{h_k }^2 + {h_m }^2}}{{{h_k }^2 - {h_m }^2}}} {f_j}^2{V_{jm}}^*{V_{jk }}{V_{im }}\right] \; . \label{eqn:21}
\eeq
The RGE for the squares of the absolute values of the CKM matrix elements, i.e. the rephasing invariant variables, can now be calculated, where one can easily show the following relation holds:     
\bea
16 \pi ^2 \frac{d \left| V_{ij} \right|^2}{dt} &=&S(t)\left\{ 3  \left| V_{ij} \right|^2 \left(f_i^2 + h_j^2 - \sum\limits_k f_k^2 \left| V_{kj} \right|^2 - \sum\limits_k h_k^2 \left| V_{ik} \right|^2 \right) - 3 f_i^2\sum\limits_{k \ne i} \frac{1}{f_i^2 - f_k^2} (2 h_j^2 \left| V_{kj} \right|^2 \left| V_{ij} \right|^2 + \sum\limits_{l \ne j} h_l^2 V_{iklj}) \right. \nn \\
&& \hspace{1cm} \left. - 3 h_j ^2\sum\limits_{l \ne j}  \frac{1}{h_j^2 - h_l^2} \left(2 f_i^2 \left| V_{il} \right|^2 \left| V_{ij} \right|^2 + \sum\limits_{k \ne i} f_k^2 V_{iklj} \right) \right\} \; , \label{eqn:22}
\eea
where
\beq
{V_{iklj}} = 1 - {\left| {{V_{il}}} \right|^2} - {\left| {{V_{kl}}} \right|^2} - {\left| {{V_{kj}}} \right|^2} - {\left| {{V_{ij}}} \right|^2} + {\left| {{V_{il}}} \right|^2}{\left| {{V_{kj}}} \right|^2} + {\left| {{V_{kl}}} \right|^2}{\left| {{V_{ij}}} \right|^2} \; . \label{eqn:23}
\eeq
Note that this evolution does not explicitly depend on the gauge couplings, but on the evolution of the Yukawa couplings.
 
\par The structure of the one-loop RGE for the gauge couplings is given by:\footnote{Here our calculations agree with the results of Ref.\cite{Bhattacharyya:2006ym}, since we need to double count the contributions from both the right- and left-handed KK modes for the internal fermion loop, which is in contrast with the results of Ref.\cite{Datta:2010us} and \cite{Perez-Lorenzana:2010zzi}.}  
\beq
16{\pi ^2}\frac{{d{g_i}}}{{dt}} = \left[{b_i}^{SM} + (S(t) - 1){{\tilde b}_i}\right]{g_i}^3 \; , \label{eqn:24}
\eeq
where $b_i^{SM} = \displaystyle \left( \frac{41}{10}, - \frac{19}{6}, - 7 \right)$ and $\tilde{b}_i = \displaystyle \left( \frac{81}{10}, \frac{7}{6}, - \frac{5}{2} \right)$. Due to the contributions from the KK modes, the gauge couplings are also altered to a power law evolution. Since the appearance of the extra dimensions causes their running to vary much more rapidly, it may well be possible to contemplate a scenario in which perturbative gauge coupling unification is preserved to a much lower energy scale. Therefore, if the unification scale is sufficiently low we can actually avoid the gauge hierarchy problem. 

\section{Properties of RGE Evolution}\label{sec:3}

\par From this full set of one-loop coupled RGE for the Yukawa couplings and the CKM matrix, together with those for the gauge coupling equations, one can obtain the renormalization group flow of all observables related to up- and down-quark masses and the CKM matrix elements. We assume the fundamental scale is not far from the range of LHC scale, and set the compactification radii to be $R^{-1} = 1$ TeV, 2 TeV, and 10 TeV respectively.  In the limit when the energy scale is much smaller than $R^{-1}$, since the energy of the system is less than the excitations of the first KK modes, the theory reduces to the usual 4-dimensional SM, and the existence of the KK excitations are ignored. When $\mu > R^{-1}$, excitations of many KK modes become possible, and the contributions of these KK states must be included in all physical calculations. This is characterized by the second term in Eq.(\ref{eqn:3}) in the general beta function.  Hence we apply the full RGE Eqs.(\ref{eqn:16},\ref{eqn:19},\ref{eqn:22}) to explore the scaling dependence behaviors of these physical observables.  For the gauge couplings we take the initial inputs $\alpha_1(M_Z) = 0.01696$, $\alpha_2(M_Z) = 0.03377$, and $\alpha_3(M_Z) = 0.1184$. Once the energy passes $R^{-1}$ the excited KK modes tend to increase rapidly this running of the gauge couplings, and ultimately change the scale dependence of the gauge couplings from logarithmic to those of a power law as a function of $\mu$.  Quantitatively, due to the fast running of the gauge couplings, we find they nearly meet at around $t = 5.8$, $6.5$, 8.2 (that is, for $\mu \simeq 30$, $60$, $330$ TeV) for radii $R^{-1}= 1$, 2, 10 TeV respectively.  

\par The extra dimensions naturally lead to gauge coupling unification at an intermediate mass scale. To illustrate the power law dependence of the Yukawa couplings quantitatively, we take $m_u(M_Z) = 1.27$ MeV, $m_c(M_Z) = 0.619$ GeV, $m_t (M_Z) = 171.7$ GeV, $m_d(M_Z) = 2.90$ MeV, $m_s(M_Z) = 55$ Mev , $m_b (M_Z) = 2.89$ GeV, $m_e(M_Z) = 0.48657$ MeV, $m_\mu(M_Z) = 102.718$ MeV, and $m_\tau(M_Z) = 1746.24$ MeV \cite{Xing:2007fb} for the initial input values for fermion masses, and run the RGE of the Yukawa couplings from $M_Z$ up to the GUT scale for our three different compactification radii $R^{-1} = 1$ TeV, 2TeV, and 10 TeV . The initial Yukawa couplings are given by the ratios of the fermion masses to the Higgs vacuum expectation value. 

\par
\begin{figure}[th]
\begin{center}
\epsfig{file=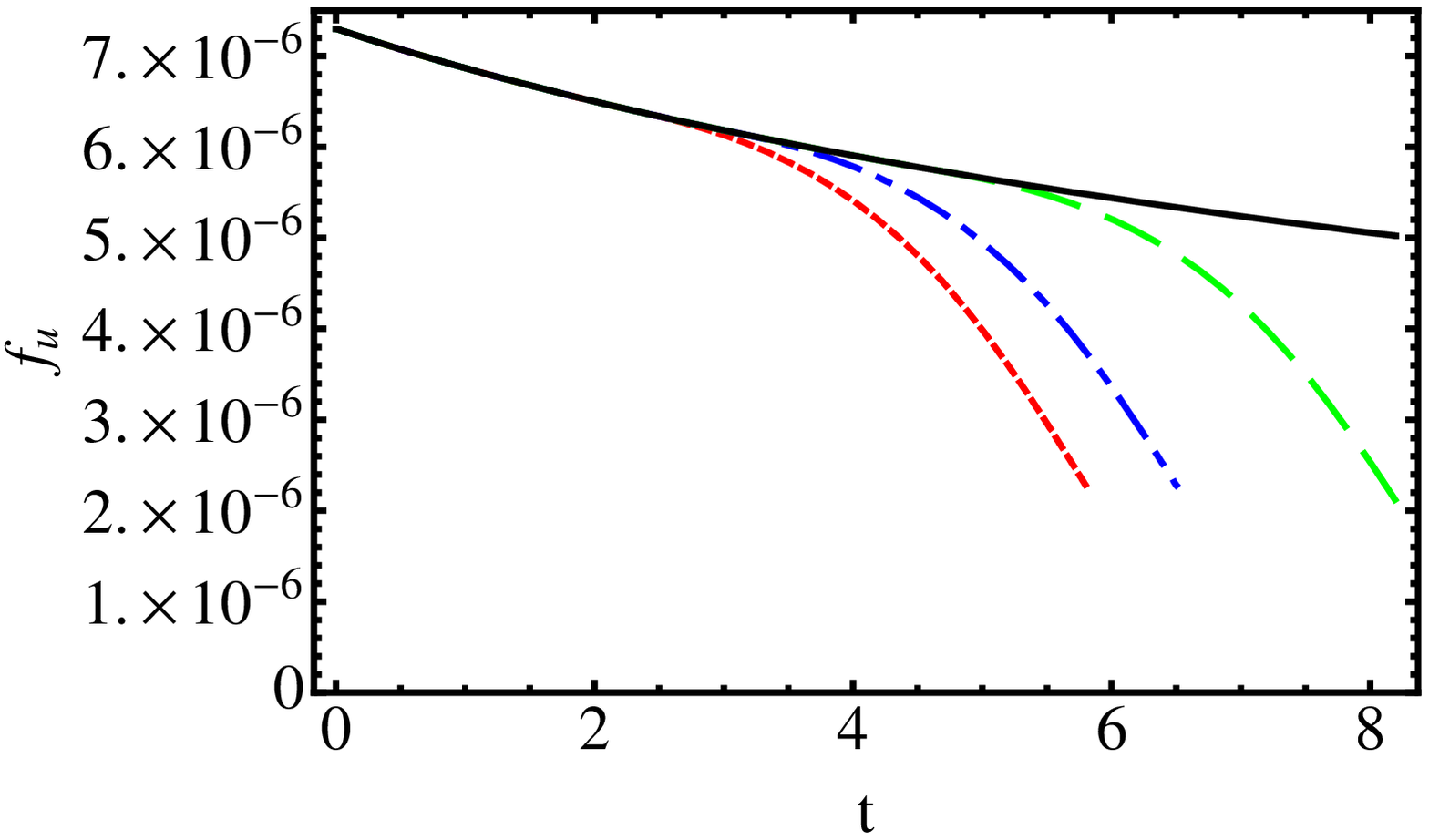,width=.4\textwidth}
\epsfig{file=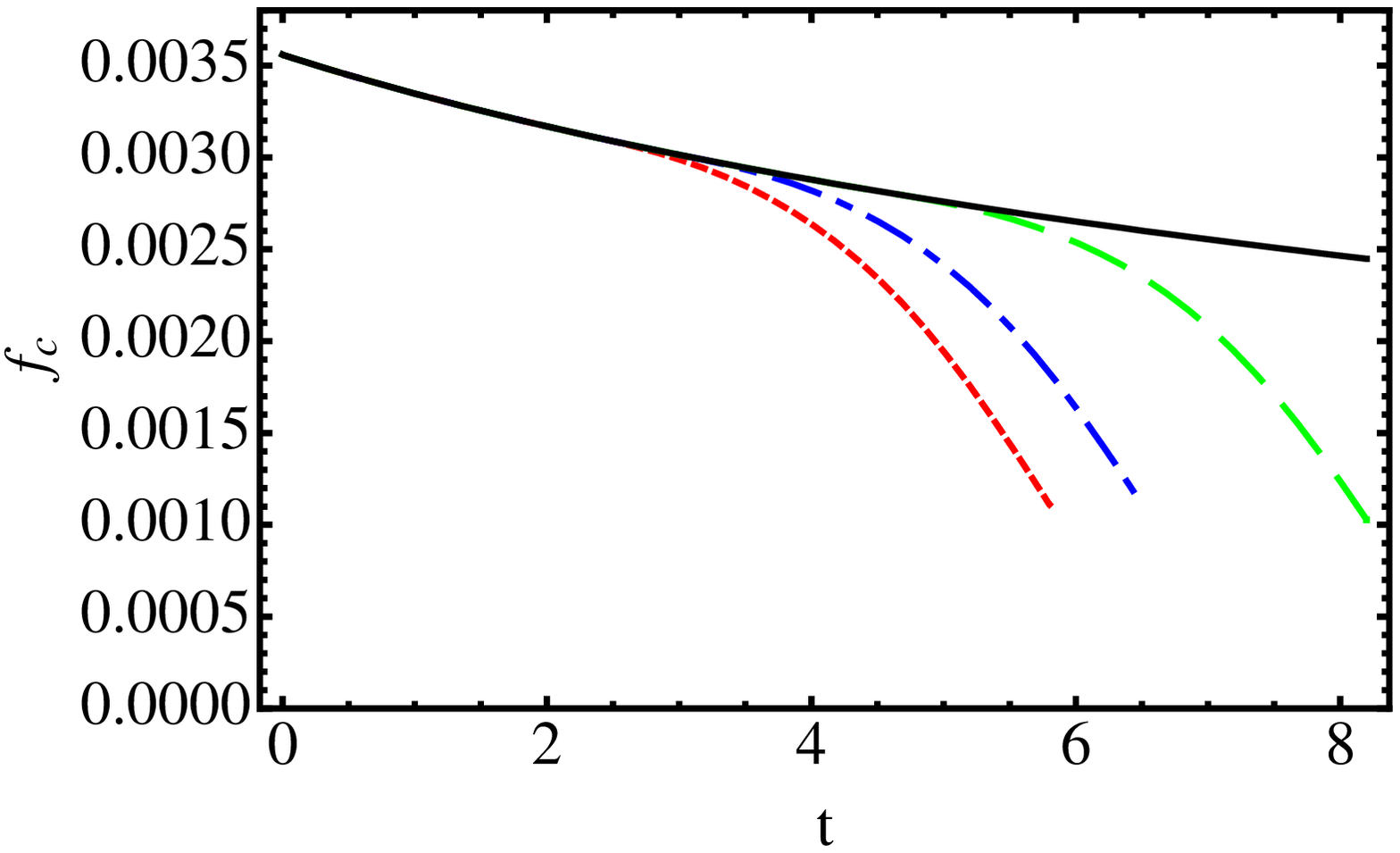,width=.4\textwidth}
\epsfig{file=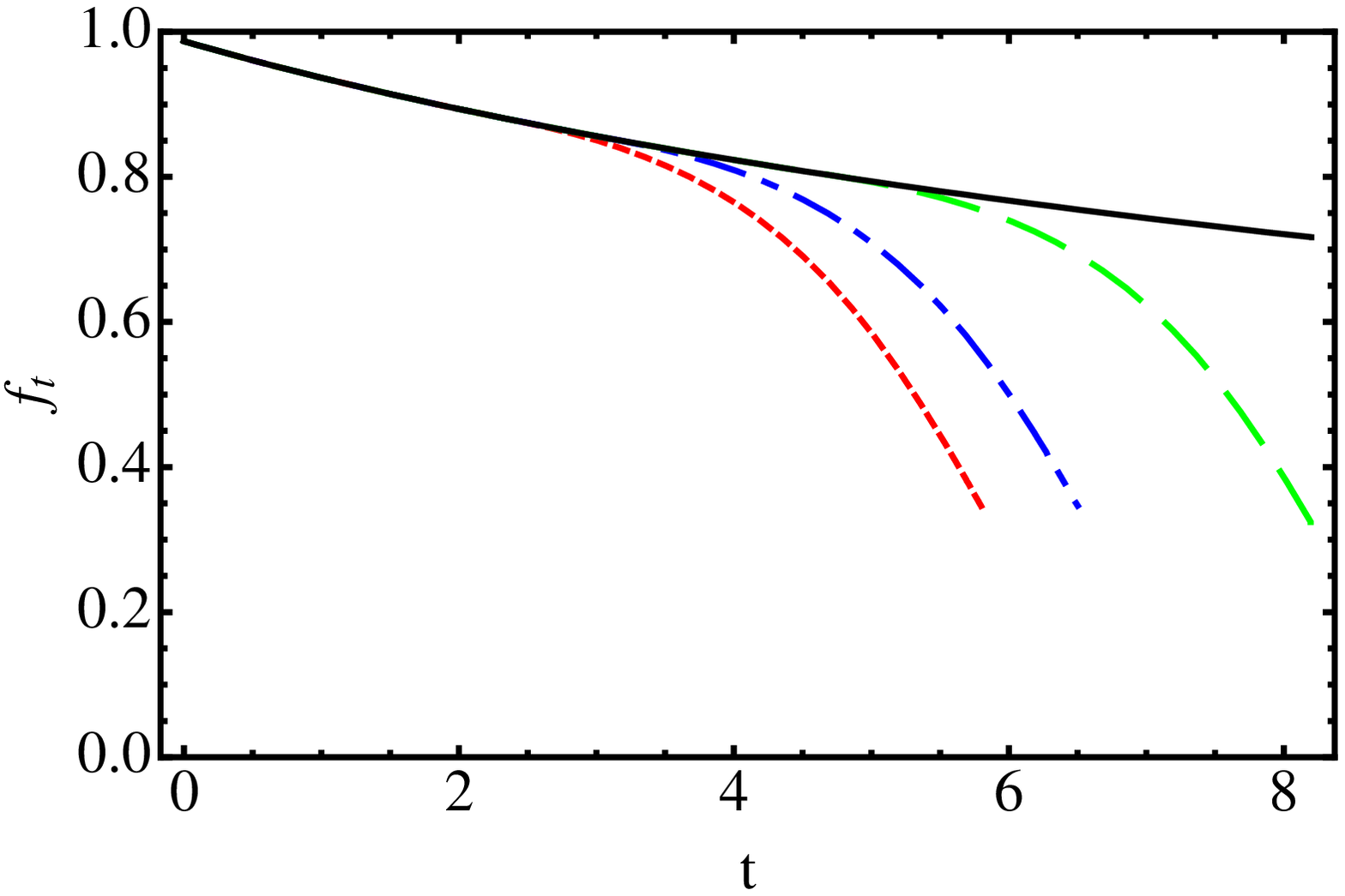,width=.4\textwidth}
\caption{\sl The evolution of the Yukawa coupling $f_i$ ($i = u$ top left panel, $c$ top right panel, $t$ bottom panel), where the solid line is the SM, the dotted line is the $R^{-1} = 1$ TeV UED case, the dotted-dashed line is the 2 TeV UED case and the dashed line is the 10 TeV UED case.}
\label{fig:3-1}
\end{center}
\end{figure}

\begin{figure}[th]
\begin{center}
\epsfig{file=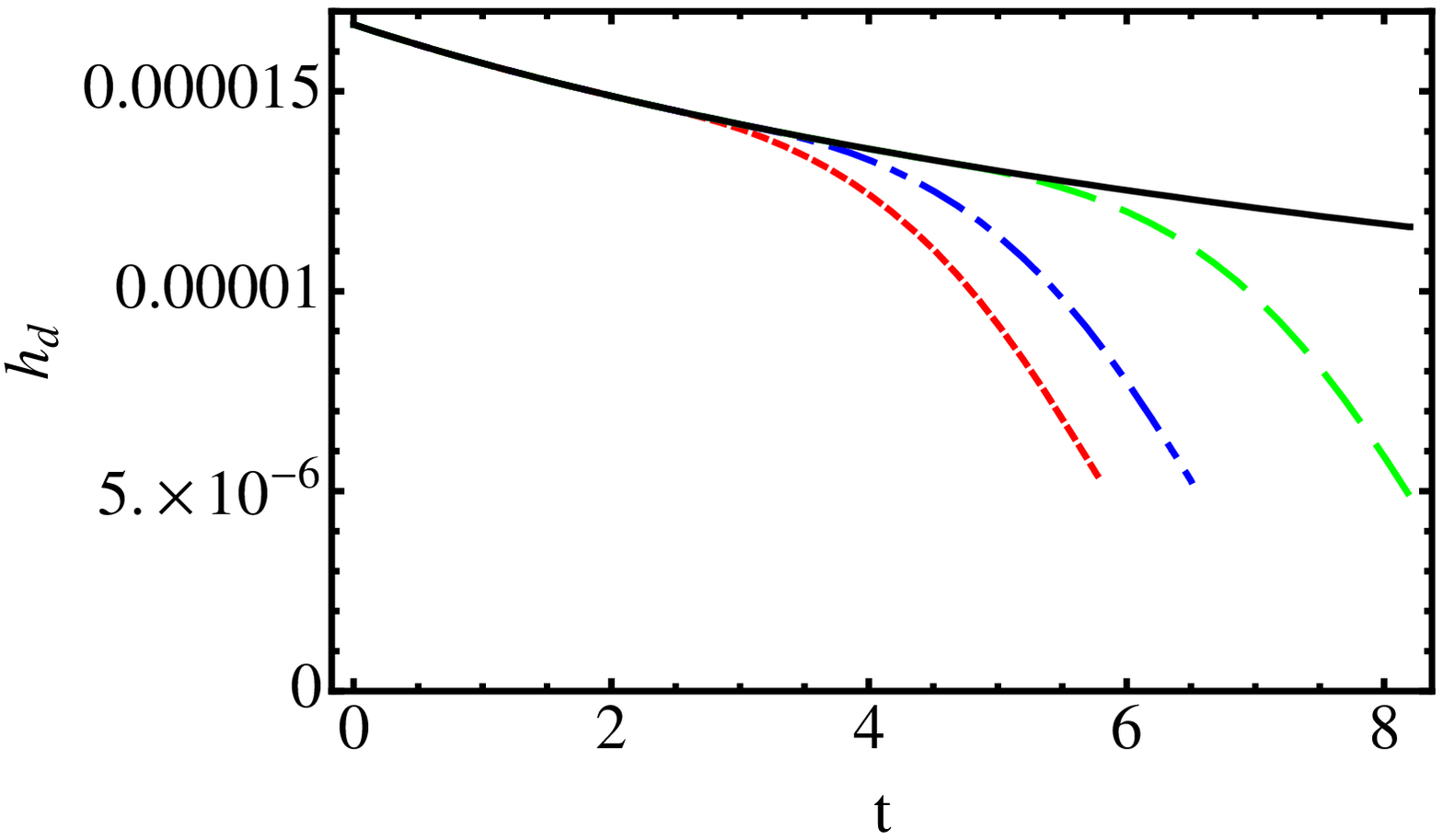,width=.4\textwidth}
\epsfig{file=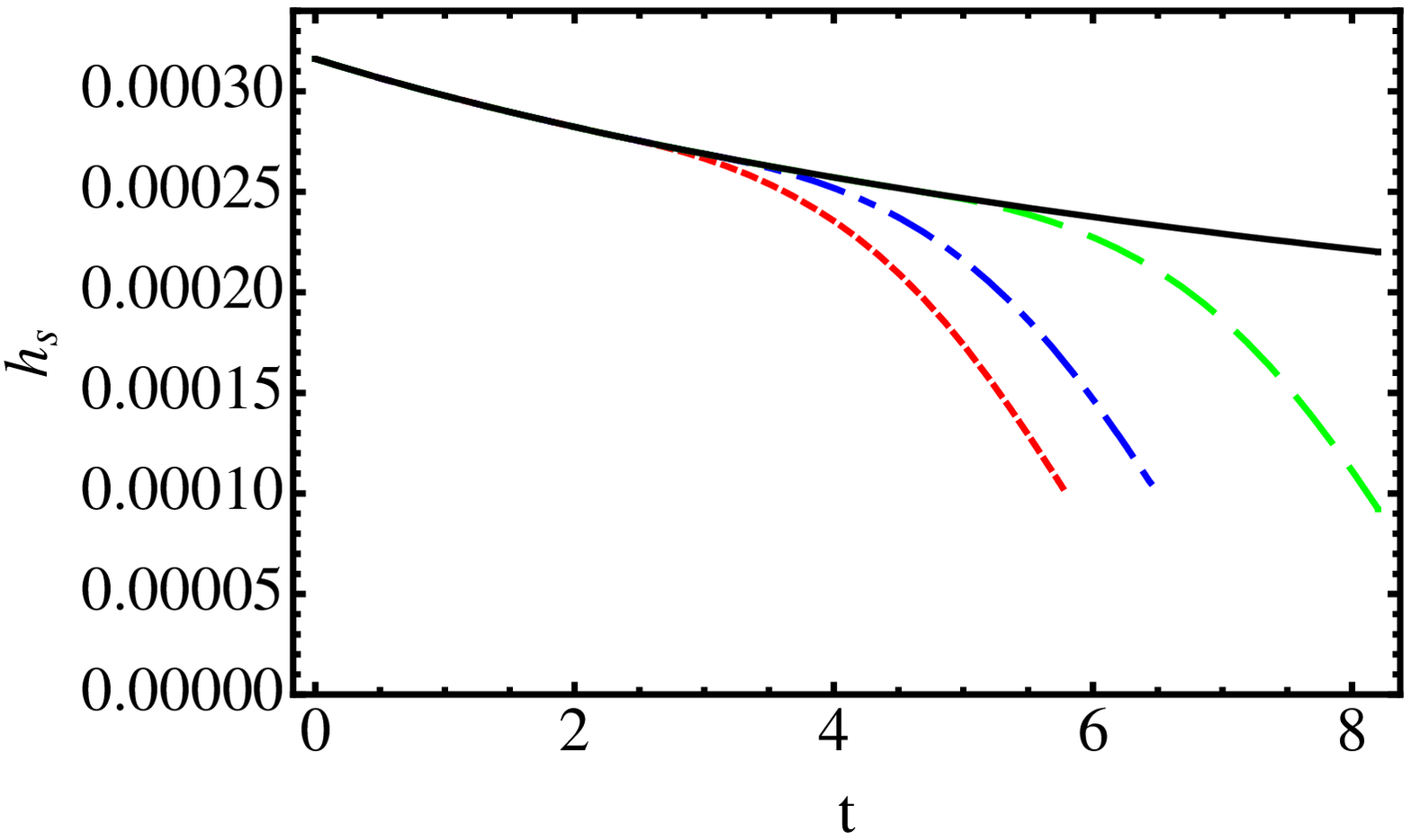,width=.4\textwidth}
\epsfig{file=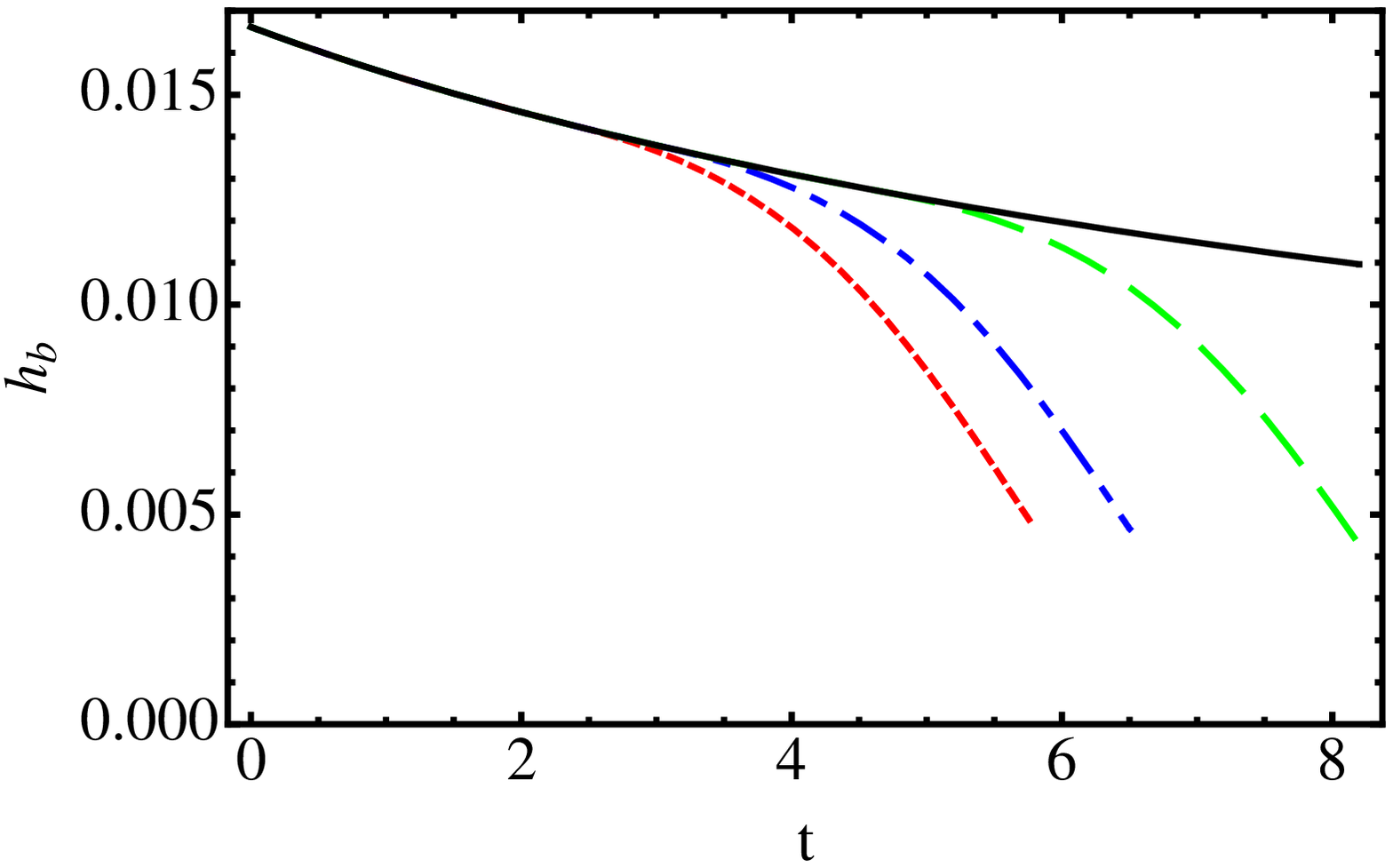,width=.4\textwidth}
\caption{\sl The evolution of the Yukawa coupling $h_j$ ($j = d$ top left panel, $s$ top right panel, $b$ bottom panel), where the solid line is the SM, the dotted line is the $R^{-1} = 1$ TeV UED case, the dotted-dashed line is the 2 TeV UED case and the dashed line is the 10 TeV UED case.}
\label{fig:3-4}
\end{center}
\end{figure}

 As illustrated in Figs.\ref{fig:3-1}-\ref{fig:3-4}, the Yukawa couplings evolve in the usual logarithmic fashion when the energy is below 1 TeV, 2 TeV, and 10 TeV for the three different cases. However, once the first KK threshold is reached, the contributions from the KK states become more and more significant. The second terms on the right hand side of Eq.(\ref{eqn:11}) depend explicitly on the cutoff $\Lambda$, which have finite one-loop corrections to the beta functions at each massive KK excitation level. Therefore, the running of the Yukawa couplings, or more precisely, the one-loop KK corrected effective four dimensional Yukawa couplings, begins to deviate from their normal orbits and start to evolve faster and faster. For the compactification radius $R^{-1} = 1$ TeV, the Yukawa couplings evolve faster than the other two, reaching its maximum value at the unification scale of around 30 TeV, after that point their evolution will ``blow-up" due to the faster running of the gauge couplings and new physics would come into play. For the radius $R^{-1} = 2$ TeV, we find similar behavior to the $R^{-1} = 1$ TeV case, where the ``blow-up" scale is not very far from that of 1 TeV case. However, for the third choice of radius, since the compactification radius is now much higher than the other two, we need more energy to push it further toward its ``blow-up" point, which is at a higher unification scale.  We also observe that the Yukawa couplings are quickly evolving to zero, however, a satisfactory unification of these seems to still be lacking. The first generation $f_u$ and $h_d$ are driven to the order of $10^{-6}$, while the $f_t$, the heaviest one, is driven to the order of $10^{-1}$. In Ref.\cite{Dienes:1998vg}, however, an additional scalar superfield is introduced which is located at the orbifold fixed points and therefore lacks a KK tower. This scalar superfield's contribution to the Higgs wavefunction renormalization causes the unification of the Yukawa couplings to become feasible. In the UED scenario, the unification of the Yukawa couplings is very desirable due to the fast power law running. This feature thus has the potential to address the problem of fermion mass hierarchy. We plan to come back to this point in a future work. As such, we have so far observed the Yukawa couplings all decrease with increasing energy, which agrees with what is observed in the SM, however, the Yukawa couplings are driven dramatically towards extremely weak values at a much faster rate. This is an interesting feature that distinguishes the UED model from that of the SM. 

%%%%%%%%%%%%%%%%
%%%%%%%%%%%%%%%%
%%%%%%%%%%%%%%%%
%\begin{figure}[th]
%\begin{center}
%\epsfig{file=fc.eps,width=.8\textwidth}
%\caption{\sl The evolution of the Yukawa coupling $f_c$, where the solid line is the SM, the dotted line is the $R^{-1} = 1$ TeV UED case, the dotted-dashed line is the 2 TeV UED case and the dashed line is the 10 TeV UED case.}
%\label{fig:3-2}
%\end{center}
%\end{figure}
%%%%%%%%%%%%%%%%
%%%%%%%%%%%%%%%%
%\begin{figure}[th]
%\begin{center}
%\epsfig{file=ft.eps,width=.8\textwidth}
%\caption{\sl The evolution of the Yukawa coupling $f_t$, where the solid line is the SM, the dotted line is the $R^{-1} = 1$ TeV UED case, the dotted-dashed line is the 2 TeV UED case and the dashed line is the 10 TeV UED case.}
%\label{fig:3-3}
%\end{center}
%\end{figure}
%%%%%%%%%%%%%%%%
%%%%%%%%%%%%%%%%
%%%%%%%%%%%%%%%%
%%%%%%%%%%%%%%%%
%\begin{figure}[th]
%\begin{center}
%\epsfig{file=hs.eps,width=.8\textwidth}
%\caption{\sl The evolution of the Yukawa coupling $h_s$, where the solid line is the SM, the dotted line is the $R^{-1} = 1$ TeV UED case, the dotted-dashed line is the 2 TeV UED case and the dashed line is the 10 TeV UED case.}
%\label{fig:3-5}
%\end{center}
%\end{figure}
%%%%%%%%%%%%%%%%
%%%%%%%%%%%%%%%%
%\begin{figure}[th]
%\begin{center}
%\epsfig{file=hb.eps,width=.8\textwidth}
%\caption{\sl The evolution of the Yukawa coupling $h_b$, where the solid line is the SM, the dotted line is the $R^{-1} = 1$ TeV UED case, the dotted-dashed line is the 2 TeV UED case and the dashed line is the 10 TeV UED case.}
%\label{fig:3-6}
%\end{center}
%\end{figure}
%%%%%%%%%%%%%%%%

\par
\begin{figure}[th]
\begin{center}
\epsfig{file=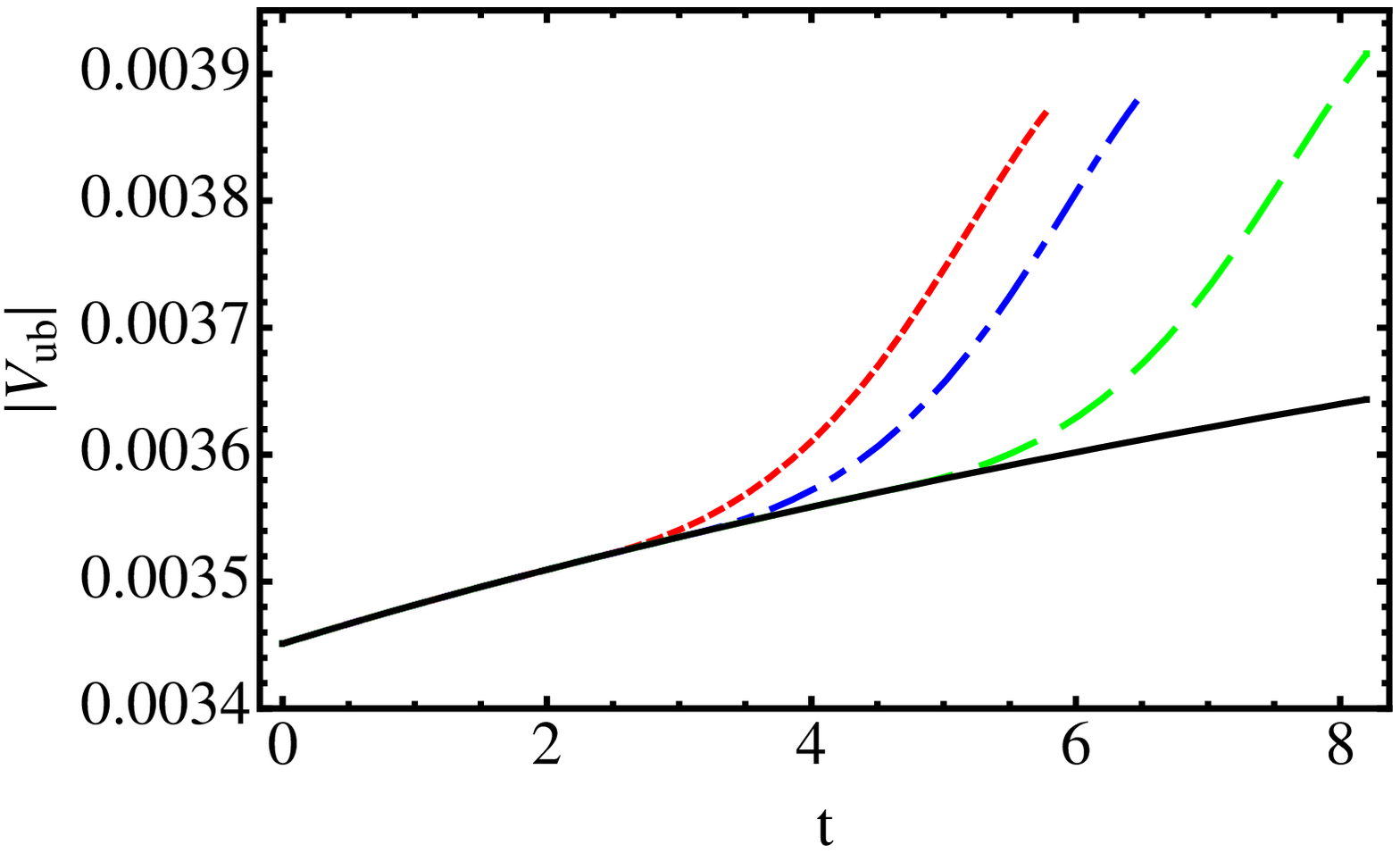,width=.4\textwidth}
\epsfig{file=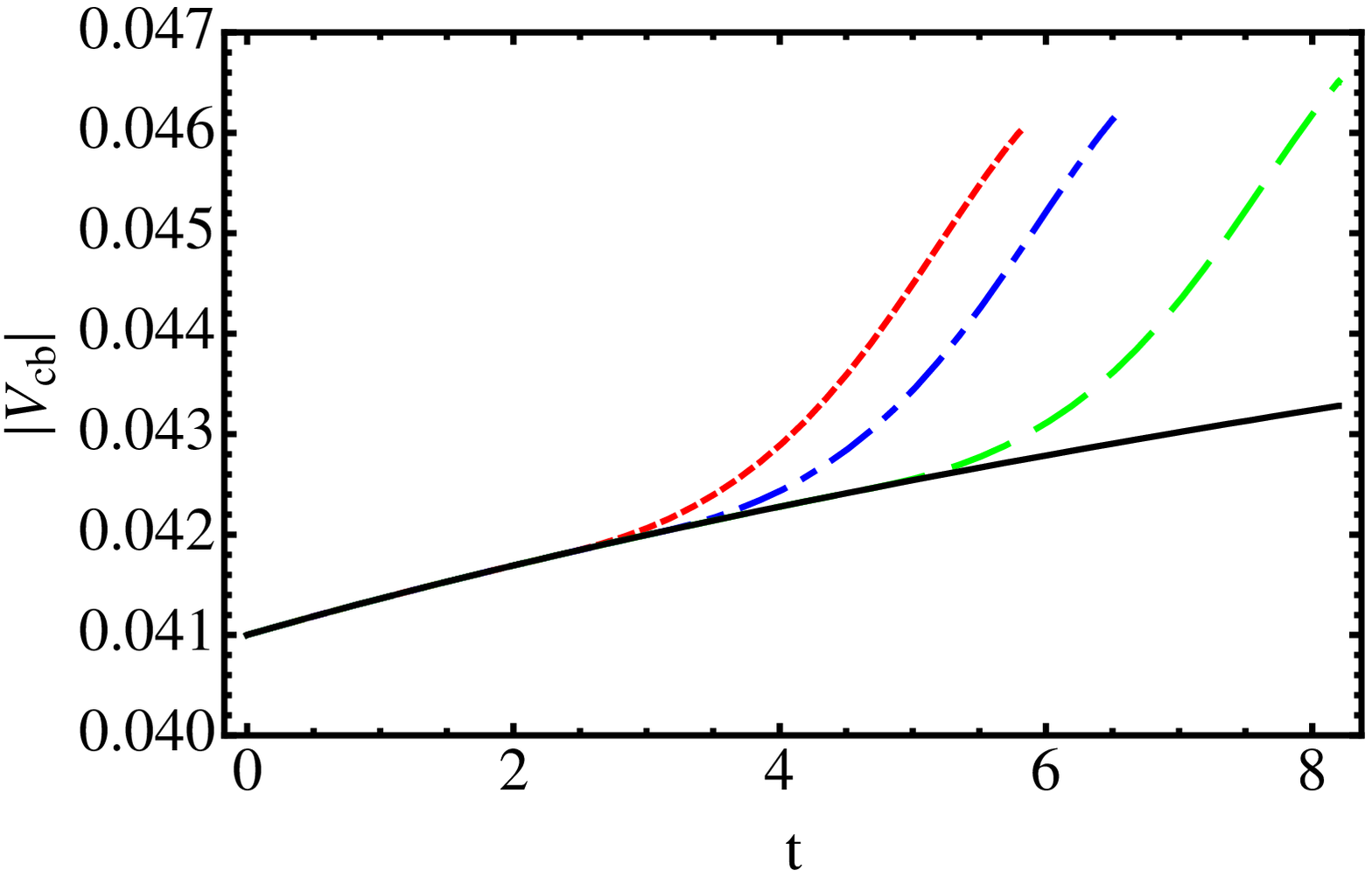,width=.4\textwidth}
\epsfig{file=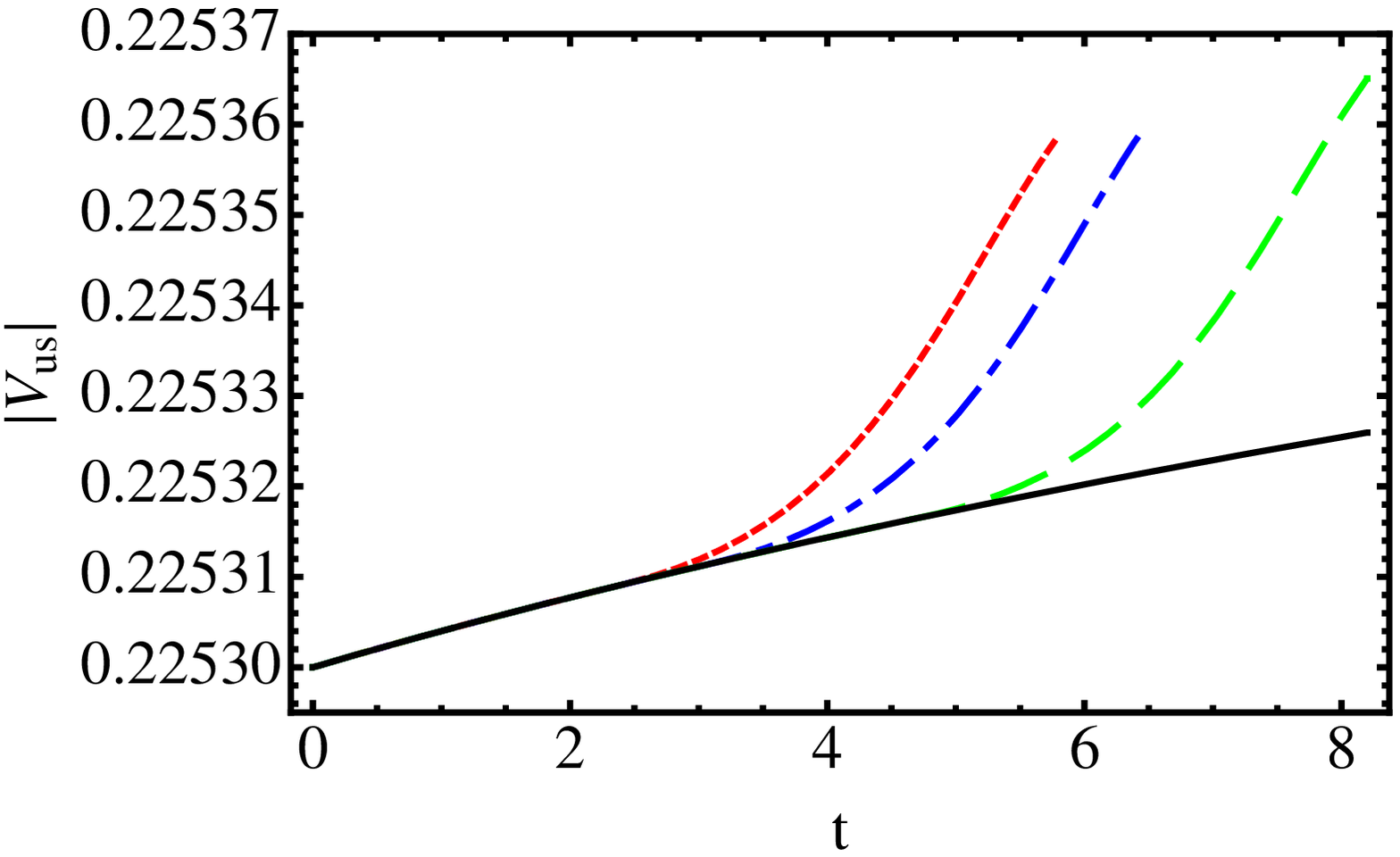,width=.4\textwidth}
\epsfig{file=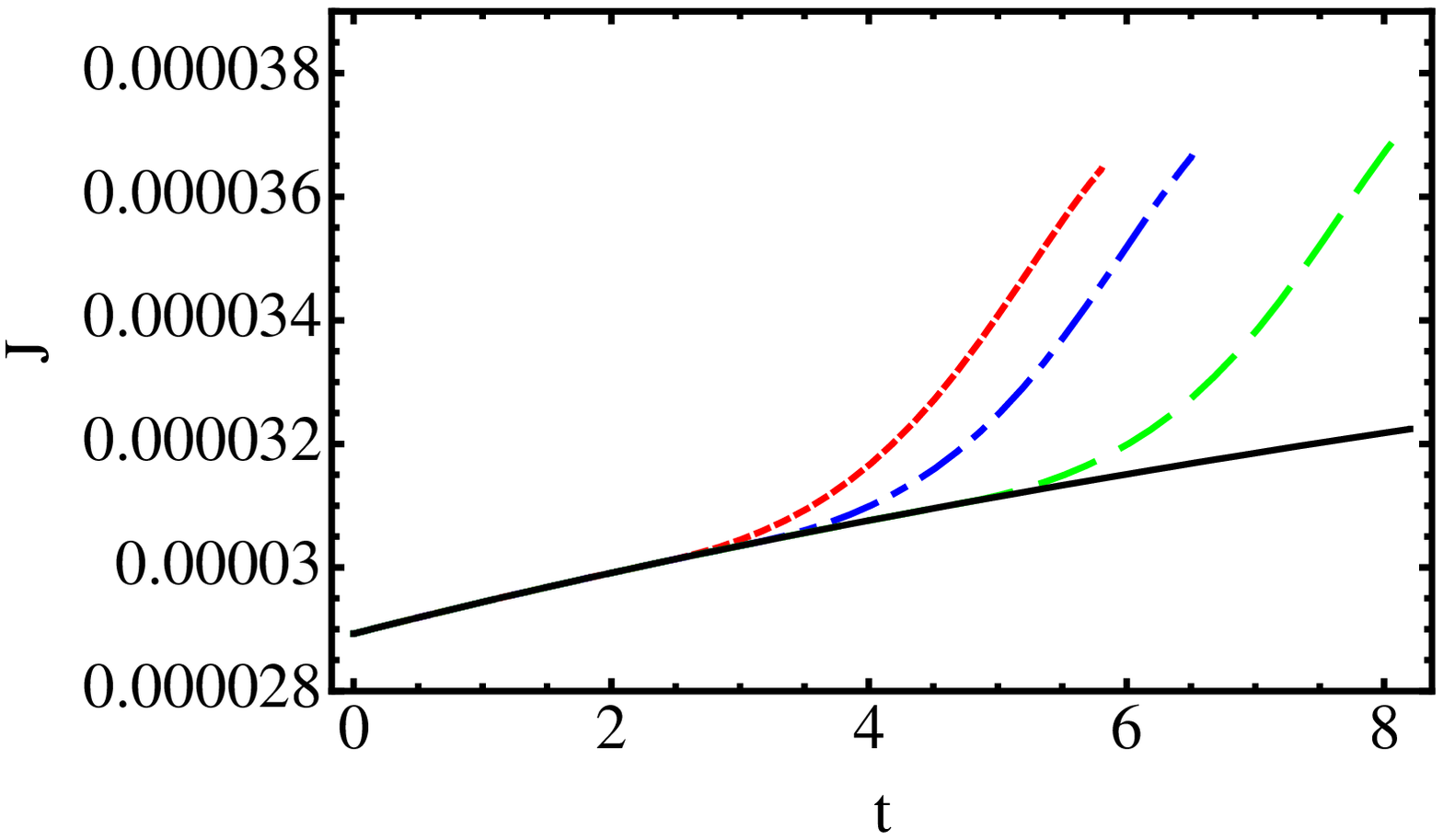,width=.4\textwidth}
\caption{\sl The evolution of the CKM matrix element $|V_{ub}|$, top left panel, $|V_{cb}|$ top right panel, $|V_{us}|$ bottom left panel, and $J$, bottom right panel; where the solid line is the SM, the dotted line is the $R^{-1} = 1$ TeV UED case, the dotted-dashed line is the 2 TeV UED case and the dashed line is the 10 TeV UED case.}
\label{fig:3-7}
\end{center}
\end{figure}

 We next turn our attention to the quark flavor mixings. Because of the arbitrariness in choice of phases of the quark fields, the phases of individual matrix elements of $V_{CKM}$ are not themselves directly observable. We therefore use the absolute values of the matrix element $|V_{ij}|$ as the independent set of rephasing invariant variables. Of the nine elements of the CKM matrix, only four of them are independent, which is consistent with the four independent variables of the standard parameterization of the CKM matrix. Among these, the complex phase of the CKM matrix characterizes CP-violating phenomena, which have been unambiguously verified in a number of $K - \bar{K}$ and $B - \bar{B}$ systems. Conventionally, one uses the Jarlskog rephasing invariant parameter $J = \mathrm{Im} V_{ud}V_{cs}V^*_{us} V^*_{cd}$ to present CP violation phenomena. Its square can be written as follows:
\beq
{J^2} = {\left| {{V_{11}}} \right|^2}{\left| {{V_{22}}} \right|^2}{\left| {{V_{12}}} \right|^2}{\left| {{V_{21}}} \right|^2} - \frac{1}{4}{\left(1 - {\left| {{V_{11}}} \right|^2} - {\left| {{V_{21}}} \right|^2} - {\left| {{V_{22}}} \right|^2} - {\left| {{V_{12}}} \right|^2} + {\left| {{V_{11}}} \right|^2}{\left| {{V_{22}}} \right|^2} + {\left| {{V_{21}}} \right|^2}{\left| {{V_{12}}} \right|^2}\right)^2} \; . \label{eqn:25}
\eeq
\noindent For definiteness, we choose the $|V_{ub}|$, $|V_{cb}|$, $|V_{us}|$ and $J$ as the four independent parameters of $V_{CKM}$, and take the initial values $|V_{ub}| = 0.00347$, $|V_{cb}| = 0.0410$, $|V_{us}| = 0.2253$ and $J = 2.91 \times 10^{-5}$ \cite{Nakamura:2010zzi}. In Fig.\ref{fig:3-7} we plot the energy dependence of these four variables from the weak scale all the way up to the unification scale for different values of compactification radii $R$. 

\par We observe from these plots the following; the CKM matrix elements $V_{ub} \simeq \theta_{13} e^{-i \delta}$, $V_{cb} \simeq \theta_{23}$, can be used to observe the mixing angles $\theta_{13}$ and $\theta_{23}$ and that they increase with the energy scale; the variation rate becoming faster once the KK threshold is passed. For the mixings related to the third family, the UED effects become sizable and the mixing angles $\theta_{13}$ and $\theta_{23}$ change at a level of 15\% between $M_Z$ and the unification scale, in contrast with the SM, in which the angles only rise by around 5\% at similar energy scales. By contrast, the variation of the Cabibbo angle appears to be the least sensitive. Due to the smallness of the Yukawa coupling terms on the right hand side of Eq.(\ref{eqn:22}), the renormalization group flow of the mixing between the first two families turns out to be very small. Although the mixing angle increases all the time, it is rather inert, even in the UED model. It has a maximum variation around the order of $\lambda^6$, for the Wolfenstein parameter $\lambda=0.22$.  However, for the parameter $J$, the characteristic parameter for the CP non-conservation effects, its variation becomes very significant. The larger the value of the compactification radius $R$, the faster $J$ evolves to reach its maximum. We observe an approximate 30\% increase for $J$ at the unification scale compared with its initial value.  The absolute values of all the remaining magnitudes of the CKM matrix elements can be obtained from the unitarity equations, thus we could determine the renormalization group evolutions of the full CKM matrix. As depicted in Fig.\ref{fig:3-7}, with increasing energy the running of the CKM matrix shows a pronounced pattern at the point where the KK modes are excited. As can be seen from Eq.(\ref{eqn:22}), the evolution of the CKM matrix is governed by the Yukawa couplings and the factor $S(t)$. They evolve faster in the region where the power law scaling of the Yukawa couplings becomes substantial. This effect is explicit for mixings involving the third family, due to the large value of their Yukawa couplings. 

%%%%%%%%%%%%%%%%
%%%%%%%%%%%%%%%%
%%%%%%%%%%%%%%%%
%\begin{figure}[th]
%\begin{center}
%\epsfig{file=Vcb.eps,width=.8\textwidth}
%\caption{\sl The evolution of the CKM matrix element $|V_{cb}|$; where the solid line is the SM, the dotted line is the $R^{-1} = 1$ TeV UED case, the dotted-dashed line is the 2 TeV UED case and the dashed line is the 10 TeV UED case.}
%\label{fig:3-8}
%\end{center}
%\end{figure}
%%%%%%%%%%%%%%%%
%%%%%%%%%%%%%%%%
%\begin{figure}[th]
%\begin{center}
%\epsfig{file=Vus.eps,width=.8\textwidth}
%\caption{\sl The evolution of the CKM matrix element $|V_{us}|$; where the solid line is the SM, the dotted line is the $R^{-1} = 1$ TeV UED case, the dotted-dashed line is the 2 TeV UED case and the dashed line is the 10 TeV UED case.}
%\label{fig:3-9}
%\end{center}
%\end{figure}
%%%%%%%%%%%%%%%%
%%%%%%%%%%%%%%%%
%\begin{figure}[th]
%\begin{center}
%\epsfig{file=J.eps,width=.8\textwidth}
%\caption{\sl The evolution of the CKM matrix element $J$; where the solid line is the SM, the dotted line is the $R^{-1} = 1$ TeV UED case, the dotted-dashed line is the 2 TeV UED case and the dashed line is the 10 TeV UED case.}
%\label{fig:3-10}
%\end{center}
%\end{figure}
%%%%%%%%%%%%%%%%

%%%%%%%%%%%%%%%%%%%%%%%%%%%%%%%%%

\section{Summary}\label{sec:4}

\par UED models with a compactification radius near the TeV scale promise to provide an exciting set of observable phenomenology for collider physics. If the compactification radius $R$ is sufficiently large, all SM particles KK partners might be detected at accelerators, such as the LHC, which can run at center of mass energies of 10 TeV. In the UED scenario we have examined the cumulative contribution of these KK states to the renormalization group evolution of the Yukawa couplings and the CKM matrix. Due to the power law running of the gauge couplings, we can bring the unification scale down to an explorable range at the LHC scale. This fact is in clear contrast to the SM. We have plotted the evolution of the Yukawa couplings in the UED for different compactification radii $R$, where as depicted in Figs.\ref{fig:3-1}-\ref{fig:3-4}, the rapid decrease of the Yukawa couplings with energy is in clear contrast to the slow logarithmic running predicted by the SM.  By permitting the Yukawa couplings to receive power law corrections in the UED model we show that extra dimensions provide a potential mechanism for explaining the fermion mass hierarchy. To introduce new fields in the bulk or at the orbifold fixed points that allows new terms in the beta function may give us an alternative evolution scenario. A more detailed analysis in this direction will be attempted in a future publication.   

\par To conclude, we have investigated the consequence of the UED model on the Yukawa couplings and CKM matrix elements evolution. We performed a qualitative study of the behavior of different mixing angles and the CP violation measure $J$ as well. The energy dependence of $|V_{us}|$ is very weak and qualitatively different from those of the mixing behaviors involving the third generation. However, when the unification scale is approached, its variation is rather rapid, in clear contrast to the slow variation of the SM. While the evolution of the Cabibbo angle is tiny, the elements $|V_{ub}|$ and $|V_{cb}|$ increase sizably, the relative deviations for these can be up to 15\% in the whole range from $M_Z$ to the GUT scale. As for the energy scaling of $J$, the contribution of KK modes is substantial. Its numerical analysis shows us that its variation can be raised to more than 30\%. The scale deviation of renormalization curves from the usual SM one depends closely on the value of the compactified radius $R$. The smaller the radius is, the higher the energy scale we need to differentiate the UED curve from the SM one.  A comparison between theoretical predictions and experimental measurements will be available once the LHC is running at its full scale and the acquisition of data becomes available. It is believed that it will set a strong limit on the parameters of this model, and a precise determination of $J$, $|V_{ub}|$ or $|V_{cb}|$ may lead to a discrimination between the SM and extra dimensional models.

%%%%%%%%%%%%%%%%%%%%%%%%%%%%%%%%%

\acknowledgments

\par The authors would like to thank R. de Mello Koch for his helpful comments and discussions. L. X. Liu would also like to thank T. K. Kuo for his support. 

%%%%%%%%%%%%%%%%%%%%%%%%%%%%%%%%%

\end{document}